\documentstyle[preprint,aps]{revtex}
\tighten
\begin{document}
\preprint{IFUM 545/FT   November 1996}
\draft

\title
 { The 2n-point renormalized coupling constants\\
in the 3d Ising model:\\
 estimates by high temperature series to order $\beta^{17}$\\}
\author{P. Butera\cite{pb} and M. Comi\cite{mc}}
\address
{Istituto Nazionale di Fisica Nucleare\\
Dipartimento di Fisica, Universit\`a di Milano\\
Via Celoria 16, 20133 Milano, Italy}
\maketitle

\begin{abstract}
Abstract: 
We compute the  2n-point renormalized coupling constants
in the symmetric phase of  the 3d Ising model  on the sc  lattice
 in terms of the high temperature expansions  $O(\beta^{17})$  
of  the Fourier transformed 2n-point connected correlation functions 
at zero momentum.

  Our high temperature estimates 
of these quantities, which enter into the small field expansion 
of the effective potential for a 3d scalar field at the IR fixed
 point or, equivalently, in the critical equation of state of the 3d
 Ising model universality class, are compared
with recent results obtained by renormalization group methods, 
strong coupling, 
 stochastic simulations as well as previous high temperature  expansions.  
\end{abstract}
 \widetext

\small
\section{Introduction}
In recent times a considerable effort has been devoted 
 to the evaluation of the 2n-point
dimensionless renormalized coupling constants (RCC's) at  zero momentum 
for the Ising model in three dimensions.
These quantities are of interest 
for  constructing the field theoretic effective 
potential\cite{jona,bender} 
of a 3d scalar field at the infrared fixed point 
or, in statistical mechanics language, for 
the formulation of the critical equation of state
of the 3d Ising 
model universality class\cite{gaunt,milo,tsypin}.
The computational methods, which so far have been  used, include 
 various approximate 
forms \cite{wetter,soko1,soko2,soko3,guida,bagnuls} 
of the renormalization group (RG), the field theoretic 
strong coupling expansion\cite{bender},
 the high temperature (HT) 
expansion\cite{gaunt,milo,essam,reisz,fisher} and
 (single- or multi-cluster) MonteCarlo 
techniques \cite{tsypin,wheater,kim,baker}.

In this note we want to discuss how helpful in getting 
a first estimate of the RCC's in 
the symmetric phase, can be   extensive 
HT expansion data published long ago\cite{km}  and so far only
 partially analyzed. 
 Indeed  expansions 
as  double series in the  HT  variables 
$v = tanh(\beta)$ and $\tau = exp(\beta H)$, 
where $\beta$ is the inverse temperature, 
are available for the Ising model free energy 
in a magnetic field $H$ on  various 2-,3- and 4-dimensional  lattices.
 In particular, in the 3d case the series extend  up to
order  $v^{17}$ for the sc lattice, up to  
$v^{13}$ for the bcc lattice and up to  
$v^{10}$ for the fcc lattice. 
 By computing the 2n-th derivative of the free energy with respect 
to the magnetic field at zero field we readily obtain
 the HT  expansion  of  the Fourier transformed 2n-point 
connected correlation function at zero momentum (also called the 2n-th
 susceptibility)

\begin{equation} 
\chi_{2n}(v)= \sum_{x_2,x_3,..x_{2n}}<s(0)s(x_2)s(x_3)..s(x_{2n})>_c
\end{equation}

  These expansions 
together with that 
of the second  moment correlation length  

$\xi^2(v) = \frac {\mu_2(v)}  {6 \chi_2(v)}$
are the essential ingredients for the calculation
 of the RCC's. 

The  expansion of the second moment of 
the 2-point correlation function $\mu_2(v)$ on the sc lattice
  has been recently 
extended  in Ref. \cite{noi}.

In terms of these quantities, the first few RCC's, 
in the symmetric phase,  
are defined\cite{bender} as the values $g^+_{2n}$, ($n\geq 2$), that
the following expressions 

\begin{eqnarray*}
g_4(v)=-\frac{V} {4!} \frac {\chi_4(v)} 
{\xi^3(v) \chi_2^2(v)}
\end{eqnarray*}

\begin{eqnarray*}
g_6(v)=\frac{V^2} {6!}(-\frac {\chi_6(v)} 
{\xi^6(v) \chi_2^3(v)} 
+ 10\frac {\chi_4^2(v)} {\xi^6(v) \chi_2^4(v)})
\end{eqnarray*}

\begin{eqnarray*}
g_8(v)=\frac{V^3} {8!}(-\frac {\chi_8(v)} 
{\xi^9(v) \chi_2^4(v)}
+56\frac {\chi_6(v)\chi_4(v)} {\xi^9(v) \chi_2^5(v)}
-280\frac {\chi_4^3(v)} {\xi^9(v) \chi_2^6(v)})
\end{eqnarray*}

\begin{eqnarray*}
g_{10}(v)=\frac{V^4} {10!}(-\frac {\chi_{10}(v)} 
{\xi^{12}(v) \chi_2^5(v)}
+120\frac {\chi_8(v)\chi_4(v)} {\xi^{12}(v) \chi_2^6(v)}
+126\frac {\chi_6^2(v)} {\xi^{12}(v) \chi_2^6(v)}\\
-4620\frac {\chi_6(v)\chi_4^2(v)} {\xi^{12}(v) \chi_2^7(v)}
+15400\frac {\chi_4^4(v)} {\xi^{12}(v) \chi_2^8(v)})
\end{eqnarray*}

take as $v \uparrow v_c$. 
The volume $V$ per lattice site has the value 1 
for the sc lattice, $4/3\sqrt{3}$ for 
the bcc lattice and $1/\sqrt{2}$
 for the fcc lattice.

 We recall that scaling implies that, as the critical temperature 
is approached from above, we have 
$\chi_{2n} \simeq  B^+_{2n} (v_c-v)^{-\gamma-(2n-2)\Delta}$, 
 where $\Delta$ is the gap exponent. If we also assume the 
 validity of hyperscaling,  we  have
$2\Delta = 3\nu+\gamma$ (where $\nu$ and $\gamma$ are 
the critical exponents of $\xi$ and $\chi$ respectively), 
so that the RCC's  are finite (and universal) quantities.
The  quantities  $g_{2n}$ are 
expected\cite{wegner} to be of the form 
$g_{2n}(v) \simeq g^+_{2n} 
+ A^+_{2n}(v_c - v)^{\theta}+...$ 
as $v \uparrow v_c$, 
where the dominant universal scaling correction exponent $\theta$ 
has the value $\theta = 0.50(2)$\cite{zinn}  for the 3d Ising model.

By  changing in the functions $g_{2n}(v)$ the  variable 
 $v$ into $y=\xi^2(v)$, we obtain the strong coupling expansions, 
through the order $y^{17}$, of the functions 
$\gamma_{2n}(y)$\cite{bender}  whose values
 at $y=\infty$ give the RCC's.

Let us add a few comments concerning the HT   
and the strong coupling 
series coefficients of the $\chi_{2n}$ on the sc lattice 
that we have tabulated, up to order 
 $v^{17}$, in the appendix 
together with the coefficients 
 of the second moment of the correlation function $\mu_2(v)$ 
   in order to provide the interested reader
 with all data we have used and thus make our calculations easily
 reproducible.
 The expansion for
$\chi_4$  was first computed \cite{gut86}
through $v^{17}$  using the data of  Ref.\cite{km},
but only recently we could check it 
against a completely independent 
 linked-cluster computation through the same order \cite{noi}.
We should only draw attention to  a minor misprint 
 in the last 2 digits of the coefficient of $v^{12}$ as reported
 in Ref.\cite{gut86}.
Concerning  the strong coupling expansions, 
we notice that in Ref.\cite{bender} 
$\gamma_6(y)$ has been  tabulated, for any space dimension,
 through order  $y^{11}$,  
while $\gamma_8(y)$ and $\gamma_{10}(y) $ through 
 order  $y^{7}$ only.
A further significant extension of all these
series can still be performed \cite{noi}: only then a
complete   check  against an independent computation  will be possible
for the coefficients tabulated here. 

While this work was being completed, we became aware 
 of a   related  work\cite{fisher}, 
 also devoted to the analysis of the data of Ref.\cite{km},  and where
also the  low temperature 
side of the critical region is studied. 
We decided therefore to present only the part of  our computation, 
 mainly concerning the higher RCC's, 
which was not already covered by the very 
 thorough discussion of Ref.\cite{fisher}.
In fact the availability of a longer  HT expansion 
of $\xi^2$  enables us
 to study  also individual RCC's rather than only ratios among them,
 and moreover gives access to the strong coupling expansions.

\section{Numerical results}

 We shall now present our estimates of  the first few RCC's as 
obtained from either the HT or the strong coupling expansions 
 and discuss various "biased" 
or "unbiased" numerical procedures.

In a first and straightforward approach we  estimate 
  $g^+_{2n}$ by evaluating at $v=v_c$\cite{betac,ferren,bloete} 
near diagonal Pad\`e approximants
(PA's)  of the 
quantity $f_{2n}(v) \equiv g_{2n}^{-\frac {2} {3n-3}}(v)$
 which has a Taylor   expansion in $v$. 
 This procedure is not convenient for extrapolating $g_{10}(v)$,
which changes its sign at some 
$0 < v_0 < v_c$. In this case we should
 consider instead the expression $(\frac {v} {v_c})^{6}g_{10}(v)$, 
 which also has a Taylor expansion in $v$.
 Thus, (biasing only $v_c$) we obtain the 
estimates: $\quad g^+_4 = 1.03(3),\quad g^+_6 = 1.93(8), \quad
g^+_8 = 1.53(36),\quad 
g^+_{10} =-2.0(9)$.

 Here, as in the rest of this  report, 
 our estimates are given by a suitably 
weighted average over the results   from 
the approximants using at least 14 series coefficients and
the uncertainties are 
 measured, conservatively, 
only on the basis of the spread of the
 results obtained from the highest approximants, always 
allowing  also for the  (much smaller) effects
 of the errors in $v_c$ and $\theta$. 

It should  be noticed  that 
the central estimate of $g^+_4$ obtained above 
 is slightly, but significantly 
larger than the  well established RG estimate
$g^+_4=0.988(4)$\cite{zinn}.

 This discrepancy leads us to investigate whether  
and to what extent these values  
are also affected by a "systematic" error 
due to the   non-analytic corrections 
to scaling which can spoil the 
convergence properties of the PA's.
 It has been suggested in Ref.\cite{roskies} 
that these corrections 
can be allowed for, or at least their effects can be 
significantly reduced, by performing the quadratic mapping
$v=v_c(1-\frac{ (1-z)^2} {(1-z/p)^2})$ with $p=2\sqrt{2}-1$.
 Essentially the same results are also  obtained 
by using appropriately designed
 first order differential approximants\cite{gutt} in which 
we can bias both $v_c$ and the 
scaling correction exponent $\theta$. We arrive thus at 
 our final set of estimates 

\begin{equation}
g^+_4 =0.987(4),\quad g^+_6 =1.57(10),\quad g^+_8 =0.90(10), \quad 
 g^+_{10} =-0.71(35).
\label{g*} \end{equation}

While the value of $g^+_4$ is only 
slightly lowered (and thereby closely reconciled with the 
 most accurate RG 
estimates), the central estimates of the higher 
$g^+_{2n}$ are significantly altered and the uncertainties 
 are reduced. 
Therefore it appears that our initial very simple numerical approach 
 was rather inadequate  and  moreover we  infer
     that the amplitudes $A^+_{2n}$ of the 
scaling correction terms increase  with $n$. 
 Finally,  if we notice that the uncertainties of our estimates 
 grow  rapidly with the order of the RCC's, 
it is clear why, with the presently available series,
 we have to restrict our calculations to the $g^+_{2n}$ with $n \leq 5$.

 It is interesting also  to study directly 
 other quantities such as, for instance, appropriate 
ratios of the functions $g_{2n}(v)$
 which  do not depend on $\xi^2$ and 
might  be   less sensitive to the scaling corrections, as 
a means to  understand better  the actual 
uncertainties of our numerical procedures. 
 We have therefore  considered the expression 
$T^+_1 \equiv \frac {g_6(v)} {g_4(v)^2}\vert_{v \uparrow v_c}$ 
and we have obtained the  
estimate $T^+_1 = 1.75(5)$  neglecting  
the confluent singularity and, otherwise, $T^+_1 = 1.59(5)$. 
 Analogously, we have also examined
$T^+_2 \equiv \frac {g_8(v)} {g_4(v)^3}\vert_{v \uparrow v_c}$ 
and have estimated $T^+_2=1.29(43)$ by the first method and 
  $T^+_2=0.92(13)$ by the second, while for 
$T^+_3 \equiv \frac {g_{10}(v)} {g_4(v)^4}\vert_{v \uparrow v_c}$ 
 we obtain $T^+_3=-0.7(7)$ and $T^+_3=-0.35(20)$, respectively.
All  estimates of the $T^+$ are then completely consistent with
 the corresponding separate estimates of the $g^+_{2n}$. 
Notice that the $T^+_i$ are simply related to the 
coefficients $F_i$ of the  small field expansion 
 of the "reduced effective potential" computed in 
Ref.\cite{guida} as follows: $T^+_1=96F_5,\quad$ $T^+_2=1728F_7$ and 
$T^+_3=\frac{331776} {10}F_9$. 

Let us  also recall that long ago  the sequence of universal 
amplitude combinations 
$I^+_{2r+3} \equiv \frac{\chi_2(v)^r 
\chi_{2r+4}(v)} {\chi_4^{r+1}(v)}\vert_{v \uparrow v_c}$, $ r \geq 1$, 
 which are strictly related to the $T^+_i$,   
 was introduced in Ref. \cite{watson} and, by  
using  twelve term series\cite{essam},
the first few $I^+_i$ were estimated to be
$I^+_5 = 7.73,\quad I^+_7 = 157.5,$
 and $ I^+_9 = 6180.$
 (with no indication of error).
 Our   estimates by using the direct PA procedure are 
$I^+_5 = 7.81(3),\quad 
 I^+_7 = 161.7(3),\quad  I^+_9 =6395.(21.)$,  while
if  we allow for  the scaling corrections, we  find 

\begin{equation}
I^+_5 = 7.92(7),\quad 
 I^+_7 = 165.(4.),\quad I^+_9 =6809.(120.)
\label{I*} \end{equation}

As it appears from the smaller difference between the results 
of the two kinds of numerical procedures, 
   the $T^+_i$ and especially the $I^+_i$ 
 turn out to be  less sensitive 
to the scaling corrections than the $g^+_{2n}$ and  therefore 
 we assume that they can be 
 be determined with higher relative accuracy. 
It is therefore interesting to notice that from the above estimate of
 $I^+_5$ we get the value $g^+_6 =1.62(6)$. 
Unfortunately however, at the present level of accuracy, 
 the other simple relations among the higher $T^+_i$
 and the $I^+_i$, 
like $T^+_2= \frac {12} {35}(+I^+_7-56I^+_5+280)$ etc., 
which follow from the definitions 
of the $g^+_{2n}$,    
cannot  be used
 for improving the estimates of the higher 
$T^+_i$, and therefore of the corresponding 
$g^+_{2n}$, by expressing them in terms of the 
$I^+_i$.  For instance,  
  $T^+_2$ turns out to be a small difference between 
large numbers 
 and   the uncertainty of $I^+_5$ is strongly amplified.
For similar reasons it is also not useful to start directly with the 
critical amplitudes of the $\chi_{2n}$. 

An unbiased study of the RCC's can be performed  
starting with the strong coupling expansion.  
In Ref.\cite{bender}   
an elaborate extrapolation procedure was proposed 
which involves the dependence of the series 
coefficients on the space dimensionality.
We have not  yet computed this dependence up to order $v^{17}$ 
and therefore 
 we cannot reproduce this procedure.
We can, however, try the simplest approach to  
 evaluate $\gamma_{2n}(\infty)$, which   consists in
 forming $[N+1/N]$ PA's  to the quantity
 $y\gamma_{2n}^{\frac {2} {3n-3}}(y)$ and in dividing them by $y$.
This procedure is not  very efficient and  the only reasonably stable
 results  obtained are: $ g^+_4=1.1(1)$,$\quad g^+_6=2.1(2)$,
$\quad g^+_8=1.9(2)$. 

We can also evaluate the ratios $T^+_i$ 
by diagonal PA's: again we find reasonable results only for 
$\quad T^+_1 = 1.81(4)$.  All these values   
 are consistent with our previous first 
evaluation of these quantities.

Alternatively, we can  generalize 
a technique introduced in Ref.\cite{nicksha}, 
 which consists in  inverting the functions
$z_{2n}=\gamma_{2n}^{-\frac {2} {3n-3}}(y)$ (after checking that the
dependence of  $z_{2n}$ on $y$ is monotonic) 
and in determining $g^+_{2n}$ from the value of $z_{2n}$
 where $y=y(z_{2n})$ diverges.
This is conveniently done by forming PA's of  
the logarithmic derivative of $y$. 
The results are then: 
$\quad g^+_4 = 1.01(2),\quad g^+_6 = 1.63(6), \quad
g^+_8 = 1.05(15)$. 
As indicated 
above, these procedures cannot be
used for computing $g^+_{10}$.

In conclusion, we believe that the general consistency among
the results obtained by applying suitable approximation 
procedures to various quantities  with somewhat different properties
 corroborates our estimates in (\ref{g*}).

\section{A comparison with other estimates}

Let us now proceed to a comparison with the  results 
 already available in the literature. 

 Our values in (\ref{g*})  for $g^+_4$ and for $g^+_6$ are  
not far from  
the estimates $g^+_4= 1.018(1)$  and  $g^+_6=1.793(16)$
 obtained in Ref.\cite{fisher} from the 
analysis of the same HT series. 
 A similar remark applies 
to the   estimates  $g^+_4= 0.988(60)$ and $g^+_6=1.92(24)$
 obtained  in Ref.\cite{reisz} from a sixteen term HT series.
Our   result for 
  $I^+_5 $ in (\ref{I*}) is also not far from  the recent 
estimate  $I^+_5 = 7.84(2)$ of Ref.\cite{fisher}. 

As to the strong coupling approach, we us recall that
in Ref.\cite{bender}   
the estimate  $g^+_6=1.2(1)$ was 
obtained  from an eleven term strong coupling series.

It is  also  interesting to perform 
a comparison  with the  results   obtained in 
the most extensive recent RG  study\cite{guida}, by the 
fixed dimension (FD) expansion\cite{zinn}  up to five
loop order, resummed by the Borel-Leroy technique
 combined with an appropriate conformal mapping.
The  estimate of $g^+_4$ agrees perfectly with ours, 
 the  central values  $g^+_6=1.603(6)$  and  $g^+_8=0.82(8)$ agree
 with ours within 
 $\simeq 2\%$ and $\simeq 9\%$, respectively. 
On the other hand the larger disagreement 
about the value of $g^+_{10}$ should not be taken 
too seriously because, as noted above,  
the  uncertainty which affects the calculation grows 
with the order of the RCC.
 Let us also return to a previous remark, in 
 noticing that from the estimates of the $F_i$
 in Ref.\cite{zinn} one arrives at the  values
  $ I^+_5=7.945(7)$, $ I^+_7=167.45(65)$ and $ I^+_9=6718.(81.)$ , 
  in very close agreement  with our estimates in (\ref{I*}).  
 Unfortunately,  $I^+_7$ and $I^+_9$ are actually rather insensitive 
to the values of $F_7$ and $F_9$.

 We  also ought to recall that  an independent 
 calculation in the FD scheme  gave the estimates
  $g^+_6\simeq 1.50$ in the two loop approximation\cite{soko1}, 
  $g^+_6\simeq 1.622$ at three loop order with Pad\`e-Borel
 resummation\cite{soko2}, and $g^+_6\simeq 1.596$ 
 at four loops\cite{soko3}.
 On the other hand, from a 3 loop computation, 
 values  for $g^+_8$ have been obtained\cite{soko2}  
 which range from $0.68$ to 
 $2.71$,  depending on the resummation procedure.  

The approximate truncation of  the RG flow equations 
studied in  Ref.\cite{wetter} yields $g^+_4= 1.2$ 
and $g^+_6=2.25$, which are both clearly  larger than 
our values, although the ratio $g^+_6/{g^+_4}^2= 1.56 $ 
agrees well with our  estimate.
An analogous, but lower order truncation 
of the  RG flow equations\cite{bagnuls}  
 had given $g^+_6 = 2.40$.

The $\epsilon=4-d$ expansion approach has not yet been pushed beyond 
order $\epsilon^3$. It has been used, in Ref.\cite{guida},
 to produce the 
(rather large) estimates $g^+_4=1.167,\quad$ $g^+_6=2.30(5),\quad$ 
 $g^+_8=1.24(8)\quad$ and $g^+_{10}=-1.97(12)$.
Notice however that, since also the estimate of $g^+_4$
 is unusually large, the corresponding 
values of the $T^+_i$ (or of the $F_i$)  
agree very closely with the results from the FD approach.
 The $\epsilon$ expansion of $T^+_i$ was examined 
also in Ref.\cite{soko1}, where by Pad\`e-Borel 
resummation the estimate $T^+_i=1.653$ was obtained.

Finally, we recall that the MonteCarlo simulations 
 of Ref.\cite{tsypin} indicate  $g^+_6 = 2.05(15)$, which is  
 not very far from our  estimate, while the simulations described in 
 Ref.\cite{kim} indicate the  values 
 $g^+_6 = 2.7(2)$ and $g^+_8 = 4.3(6)$,   
significantly larger  than both  the RG results and ours.

A summary of the present situation is presented in 
Table 1 which collects  our  estimates of the RCC's 
 along with  the corresponding ones 
 obtained by other methods.

\section{Conclusions}

We may conclude that, although the   various  
computational approaches do not yet agree perfectly,  
they do appear to converge to common estimates   
at least for the lowest RCC's.  
Therefore, in view of the  difficulty of these calculations, 
 we believe that 
 the present residual discrepancies   should not be overemphasized.
 The $\epsilon$ expansion is certainly still too short, and 
perhaps, even for the FD  expansions, 
 a further extension would be welcome.
 The HT series  presented here
are not yet long enough, the more
 so the higher the order of the RCC considered.
  Indeed, we might argue that, at the order $v^s$, the dominant 
contributions to the HT 
expansion of $\chi_{2n}(v)$ come from  
correlation functions of spins whose 
average relative distance is $\simeq s/2n$, so that present HT
expansions, in some sense,  still describe a rather "small" system.
 Analogous  problems of size also  occur 
in stochastic simulations\cite{tsypin,kim,baker}.
 Therefore  further effort would still be welcome
 to improve the 
 reliability,  the precision and, as a result, the consistency 
 of the various approaches.

\acknowledgments 
This work has been partially supported by MURST. 
We are grateful to Prof. A. 
 I. Sokolov for a very stimulating correspondence. 
We also thank to Prof. A. 
 I. Sokolov, Dr. R. Guida and Prof. J. Zinn-Justin 
 for making their results 
available to us before publication.
 
\appendix
\section{Series Expansions}

 In the case of the sc  lattice the HT expansion of the 
 susceptibilities $\chi_{2n}$ are
\scriptsize
\[\chi_2(v)= 1+ 6 v + 30 v^{ 2}+150 v^{ 3}+ 726 v^{ 4}+ 3510 v^{ 5}
+ 16710 v^{ 6}+ 79494 v^{ 7}+375174 v^{ 8} +1769686 v^{ 9}\]\[
 +8306862 v^{10}
+38975286 v^{11}+182265822 v^{12}+852063558 v^{13}+ 3973784886 v^{14}\]
\[+ 18527532310 v^{15} +86228667894 v^{16}+ 401225368086 v^{17}... \]

\[\chi_4(v)= -2  -48 v -636 v^{ 2}-6480 v^{ 3}-56316 v^{ 4}
 -441360 v^{ 5}-3208812 v^{ 6}-22059120 v^{ 7}\] \[-145118844 v^{ 8}
 -921726704 v^{ 9} -5687262012 v^{10}-34255147920 v^{11}
-202130397708 v^{12}\]
\[ -1171902072144 v^{13} -6691059944460 v^{14}-37693869995312 v^{15}
 -209838929195580 v^{16}\] \[-1155875574355632 v^{17}...\]

\FL
\[\chi_6(v)= 16+816 v+19920 v^{ 2}+336720 v^{ 3}+ 4518816 v^{ 4}
+ 51745680 v^{ 5}+ 527187600 v^{ 6}\] \[+ 4909918704 v^{ 7}
+ 42581232864 v^{ 8}+ 348466330096 v^{ 9}+ 2717492365392 v^{10}\] 
\[+20347129869456 v^{11} + 147133138147872 v^{12} 
+ 1032333377642448 v^{13}\]
\[+ 7054626581880336 v^{14} + 47100223055946160 v^{15}
+ 308027458769860704 v^{16}\] \[ + 1977507018022916016 v^{17}....\]

\FL
\[\chi_8(v)= -272 -23808 v - 917376 v^{2} 
- 23013120 v^{3} - 437798496 v^{ 4}
  - 6852038400 v^{ 5}\]\[ - 92654596992 v^{ 6} - 1117875129600 v^{ 7}
  - 12306018523104 v^{ 8}  - 125633562017024 v^{ 9}\]\[
    - 1204105704419712 v^{10} - 10936791715557120 v^{11}
 -94844317893543648 v^{12} \]\[-789993027282411264 v^{13}
 -6351007395937161600 v^{14} -49478915100503151872 v^{15}\]\[
 -374818460005448106720 v^{16} -2768750733973561834752 v^{17}...\]

\FL
\[\chi_{10}(v)= +  7936 + 1061376 v + 59036160 v^{ 2}+ 2049776640 v^{ 3}+
 52252083456 v^{ 4}\]\[ +  1067338759680 v^{ 5}+ 18429925693440 v^{ 6}+
 278749670360064 v^{ 7}+ 3786553881275904 v^{ 8}\]\[
+ 47053476826003456 v^{ 9} +
 542381843641961472 v^{10}+ 5862580439606155776 v^{11}\]\[+
 59934902216969609472 v^{12} +  583578982058859276288 v^{13}+
 5442873762995091611136 v^{14}\]\[ + 48857090955221240911360 v^{15} 
+ 423771319439035687985664 v^{16}\]\[ 
+ 3563795335882672497655296 v^{17} ...\]

\normalsize
The HT expansion of the second moment of the correlation function
  $\mu_2$ is
\scriptsize
 \[\mu_2(v)= 6v +72v^2 +582v^3 +4032v^4 +25542v^5 +153000v^6 +880422v^7    
+4920576v^8\] \[+26879670v^9 +144230088v^{10} +762587910v^{11}    
+3983525952v^{12} +20595680694v^{13} \] \[ +105558845736v^{14}
 +536926539990v^{15} +2713148048256v^{16} +13630071574614v^{17}..\]

\normalsize
The strong coupling expansions
 of the $\gamma_{2n}$ to order $y^{17}$ are
\scriptsize
\[\gamma_4(y)= \frac {y^{-3/2}} {12} \big [ 1+ 12 y 
+ 6 y^{ 2}+ 48 y^{ 3} -630 y^{ 4}+
 7272 y^{ 5} -83292 y^{ 6}+ 957312 y^{ 7}- 11035662 y^{ 8}\]
\[ +127433528 y^{ 9} -1472947908 y^{10}+ 17036529504 y^{11}
 -197169806676 y^{12}+2283416559216 y^{13}\]\[ -26463582511368 y^{14}+
 306946999598144 y^{15}-3563327123879550 y^{16}\]\[
+ 41404188226284120 y^{17}..\big] \]

\FL
\[\gamma_6(y)= \frac {y^{-3}} {30}  \big [ 1+ 18 y 
+90 y^{ 2}+ 48 y^{ 3}+ 576 y^{ 4}
 -8352 y^{ 5}
+ 114528 y^{ 6} -1528416 y^{ 7}\]\[ + 19952712 y^{ 8} -255983472 y^{ 9}+
 3240722592 y^{10} -40613845392 y^{11}+ 505052958336 y^{12}\]\[
 -6242882909472 y^{13}+ 76802505994224 y^{14} -941288338072752 y^{15}+
 11501158664782008 y^{16}\]\[ -140176233789711696 y^{17}...\big] \]

\FL

\[\gamma_8(y)=\frac {y^{-9/2}} {56}\big [1 + 24 y 
+ 192 y^{ 2}+ 576 y^{ 3} +54 y^{ 4} +6720 y^{ 5} 
-113016 y^{ 6}+ 1753632 y^{ 7}\]
\[ -25771326 y^{ 8} +
 364798032 y^{ 9} -5028161232 y^{10}+ 67958735808 y^{11}
 -904828659212 y^{12}\]\[ +
 11905472505792 y^{13} -155154712361520 y^{14}+  2006059450196288 y^{15}\]
\[ -25765180820314374 y^{16} + 329050927608994224 y^{17}...\big ] \]

\FL
\[\gamma_{10}(y)=\frac {y^{-6}} {90} \big[ 1+ 30 y+ 330 y^{ 2}
+ 1620 y^{ 3}+ 3330 y^{ 4}
  -1080 y^{ 5}+ 67200 y^{ 6} -1314720 y^{ 7}\]\[ + 22683000 y^{ 8}
  -363847600 y^{ 9}+  5564033040 y^{10} -82249187520 y^{11}+
  1185208196160 y^{12}\]\[ -16740515134800 y^{13}+  232658153938560 y^{14}
  -3190497478487440 y^{15} \]\[ + 43262377733737920 y^{16}
             -581022341984542560 y^{17}... \big] \] 


\widetext
\squeezetable
\begin{table}
\caption{A summary of the estimates of  $g^+_{2n}$ by various methods.}
\begin{tabular}{ccccc}
Method and Ref. & $g^+_4$ & $g^+_6$&  $g^+_8$& $g^+_{10}$\\
\hline
HT  &0.987(4) &1.57(10)  &0.90(10)& -0.71(35) \\
Strong coupl. &1.01(1)&1.63(5)&1.05(9)& \\
HT \cite{fisher}&1.019(6) &1.791(38)& & \\
HT \cite{reisz}&0.988(60)&1.92(24)& & \\
RG FD-expans.\cite{guida}&0.987(2)&1.603(6)&0.83(8)&-1.96(1.26) \\ 
RG FD-expans.\cite{soko2,soko3}& &1.596 & 0.68 - 2.71& \\
RG $\epsilon$-expans.\cite{guida}&1.167&2.30(5)&1.24(8)&-1.97(12)\\
RG approx.\cite{wetter}&1.2&2.25& & \\
RG approx.\cite{bagnuls}&&2.40& & \\
Strong coupl.\cite{bender} &0.986(10)&1.2(1)& & \\
MC\cite{tsypin}&0.97(2)&2.05(15)& &  \\
MC\cite{kim}&1.02&2.7(2)&4.3(6) &  \\
\end{tabular}
\end{table}

\begin{references}
\bibitem{betac} At the level of accuracy of our 
calculation equivalent results are obtained by using either 
the value $v_c=0.2180992(26)$ of Ref.\cite{ferren} or 
the more recent value $v_c=0.2180943(4)$ of Ref.\cite{bloete} used 
in Ref.\cite{fisher} with $\theta=0.54(3)$.
\end{references}
\end{document}